\documentclass[runningheads]{llncs}

\usepackage[hyphens]{url}

\usepackage{enumitem}
\usepackage{amssymb}

\usepackage[]{xcolor}
\usepackage{graphicx}
\usepackage{booktabs}
\usepackage{adjustbox}
\usepackage{hyperref}

\usepackage[%
  square,        
  comma,         
  numbers,       
  sort&compress, 
]{natbib}

\begin{document}
\title{Comparative Analysis of Engagement, Themes, and Causality of Ukraine-Related Debunks and Disinformation}
\titlerunning{Comparative Analysis of Ukraine-Related Debunks and Disinformation}
%

\author{Iknoor Singh\orcidID{0000-0002-3788-3295} \and Kalina Bontcheva\orcidID{0000-0001-6152-9600} \and Xingyi Song\orcidID{0000-0002-4188-6974} \and Carolina Scarton\orcidID{0000-0002-0103-4072}}
\authorrunning{I. Singh et al.}

%

\institute{Department of Computer Science, University of Sheffield, Sheffield (UK) \\
\email{\{i.singh, k.bontcheva, x.song, c.scarton\}@sheffield.ac.uk}}

%
\maketitle              

\begin{abstract}
This paper compares quantitatively the spread of Ukraine-related disinformation and its corresponding debunks, first by considering re-tweets, replies, and favourites, which demonstrate that despite platform efforts Ukraine-related disinformation is still spreading wider than its debunks. Next, bidirectional post-hoc analysis is carried out using Granger causality tests, impulse response analysis and forecast error variance decomposition, which demonstrate that the spread of debunks has a positive impact on reducing Ukraine-related disinformation eventually, albeit not instantly. Lastly, the paper investigates the dominant themes in Ukraine-related disinformation and their spatiotemporal distribution. With respect to debunks, we also establish that around 18\%  of fact-checks are debunking claims which have already been fact-checked in another language. The latter finding highlights an opportunity for better collaboration between fact-checkers, so they can benefit from and amplify each other's debunks through translation, citation, and early publication online. 

\end{abstract}

\section{Introduction}

Following on from and interleaved with the COVID-19 infodemic, the war in Ukraine has unleashed a new large stream of mis- and disinformation \cite{aguerri2022fight}, as evidenced, amongst others, by fact-checkers from the European Digital Media Observatory (EDMO) who found a record-high Ukraine-related disinformation in March 2022\footnote{\url{https://edmo.eu/fact-checking-briefs/}}. Examples include viral decontextualised videos from past\footnote{\url{https://www.politifact.com/factchecks/2022/may/10/facebook-posts/no-not-footage-ukraine-shooting-down-russian-plane/}} and a popular pro-Kremlin false narrative about the existence of a biolab in Ukraine funded by Joe Biden's son\footnote{\url{https://www.politifact.com/article/2022/apr/01/facts-behind-russian-right-wing-narratives-claimin/}}. 
To counter this fast-flowing disinformation, the International Fact-checking Network (IFCN) fact-checkers are working together to maintain and publish a unified database of debunks of Ukraine-related disinformation\footnote{\url{https://ukrainefacts.org/}}. 
In order to measure the effectiveness of these efforts, we carry out a comparative analysis of engagement, themes, and predictive causality of Ukraine-related debunks and disinformation. 

The novel contributions of this paper are in answering the following three key research questions through a quantitative analysis of Ukraine-related disinformation and debunks on Twitter:

\begin{itemize}[noitemsep,leftmargin=9.5mm]
    \item[\textbf{RQ1}]{What is the overall engagement of Ukraine-related disinformation and debunks on Twitter (Section \ref{engage})? }
    \item[\textbf{RQ2}]{Does the spread of debunks have a positive impact in reducing Ukraine-related disinformation (Section \ref{posthoc})?}
    \item[\textbf{RQ3}]{What are the underlying themes in Ukraine-related disinformation and their spatiotemporal characteristics on Twitter (Section \ref{topic})?}
\end{itemize} 

In the following sections, we will discuss first related work (Section \ref{background}) and then detail the data acquisition methodology for this study (Section \ref{dataset}). 

\section{Related Work}
\label{background}

Ukraine-related pro-Kremlin disinformation \cite{yablokov2022russian} is not new  \cite{mejias2017disinformation, aguerri2022fight, lange2015strategic}. For instance, \citet{lange2015strategic} and \citet{mejias2017disinformation} studied the spread of disinformation on social media after the 2014 annexation of Crimea by the Russian Federation, while \citet{erlich2021pro} investigated if Ukrainian citizens are able to discern between factual information and pro-Kremlin disinformation. Another study \cite{gerber2016does} investigated the effectiveness of Russian propaganda in swaying the views of its readers. Recently,
\citet{park2022voynaslov} released a Ukraine-related dataset of tweets and carried out an analysis of public reactions to tweets by state-affiliated and independent media. \citet{miller_inskip_marsh_arcostanzo_weir_2022} studied the spread of tweets related to hashtags that were trending in February 2022. Nonetheless, these studies do not focus specifically on comparing Ukraine-related disinformation and debunks in terms of engagement, inter-relationship, and topics. 

Prior literature on the spread of true and false information is extensive \cite{vosoughi2018spread, grinberg2019fake, shao2018spread}. 
Nevertheless, this paper is related to prior work that studied the spread and dynamics of false information and debunks on Twitter \cite{burel2020co, burel2021demographics, park2021experimental, allcott2017social, singh2021false, jiang2021reciprocal, swire2017processing, nyhan2015estimating, barrera2020facts,zhang2022investigation, recuero2022bolsonaro}.
In particular, \citet{burel2020co} compared COVID-related misinformation and fact-checks using impulse response modelling, causal analysis, and spread variance analysis, while \citet{chen2021citizens} investigated the reasons why people share fact-checks and ways to encourage this further. Also, \citet{siwakoti2021covid} showed that user engagement with fact-checks increased significantly as a result of the COVID-19 pandemic. 
However, to the best of our knowledge, no study has examined the predictive causality between Ukraine-related disinformation and debunks, or their spatiotemporal characteristics and top disinformation themes.  

\begin{table}[!t]
\small
\centering
\caption{Top domains of disinformation and debunk links.}

\scalebox{0.8}{
\begin{tabular}{p{5cm}p{5cm}}
\toprule
\textbf{Disinformation Domains} & \textbf{Debunk Domains}    \\ \midrule
facebook.com (30\%)           & dpa-factchecking.com(25\%)      \\
tiktok.com (3\%)                    & euvsdisinfo.eu (25\%)            \\
twitter.com (3\%)                & rumorscanner.com (9\%)          \\
oroszhirek.hu (2\%)                   & politifact.com (8\%)        \\
sputniknews.com (2\%)                 & factly.in  (8\%)                \\
nabd.com (2\%)            & verify-sy.com (6\%)             \\
arabic.rt.com (1\%)                      & factcrescendo.com (4\%)\\
fb.watch (1\%)                 & verafiles.org (3\%)             \\
de.news-front.info (1\%)                     & factcheck.org (2\%)         \\
Other (55\%)                         & Other (10\%)      \\
\bottomrule              
\end{tabular}
}
\label{tab:domain}
\end{table}

\section{Data}
\label{dataset}

The data underpining our analyses spans disinformation and debunks posted between 1 February and 30 April 2022. Specifically, we focus on Ukraine-related debunks and accompanying links to the corresponding disinformation 
encompassing: \emph{(i)} 110 debunks and 311 links to disinformation published by EUvsDsinfo\footnote{\url{https://euvsdisinfo.eu/}}, which primarily fact-checks pro-Kremlin disinformation; \emph{(ii)} 344 debunks  indexed by Google in the ClaimReview format
\footnote{\url{https://www.datacommons.org/factcheck/download}}, which refer to 439 disinformation links. 
See Appendix \ref{appendix:datacollect} for details on how we collect the disinformation links from debunks.
Similar to \citet{burel2020co}, in addition to the above date restrictions, we also applied keyword-based filtering\footnote{Where debunked claims were in languages other than English, these were translated automatically with Google Translate first, prior to filtering with the keywords listed here: \url{https://gist.github.com/greenwoodma/430d9443920a589b6802070f2ca54134}} in order to select only Ukraine-related debunks and disinformation.

In total, this study analyses 454 debunk URLs and 750 links to Ukraine-related disinformation. The latter are provided  by the fact-checking organisations themselves within the published debunks (see Appendix \ref{appendix:datacollect}), therefore we consider them as accurate. Table \ref{tab:domain} shows the top domains that occur within the disinformation and debunk links. The former point either to content on social media platforms or to Kremlin-backed outlets. For debunks, the main domains are EUvsDisinfo (25\%) and Dpa-factchecking (25\%).

\begin{figure*}[!tbp]
    \centering
    \includegraphics[width=0.75\textwidth]{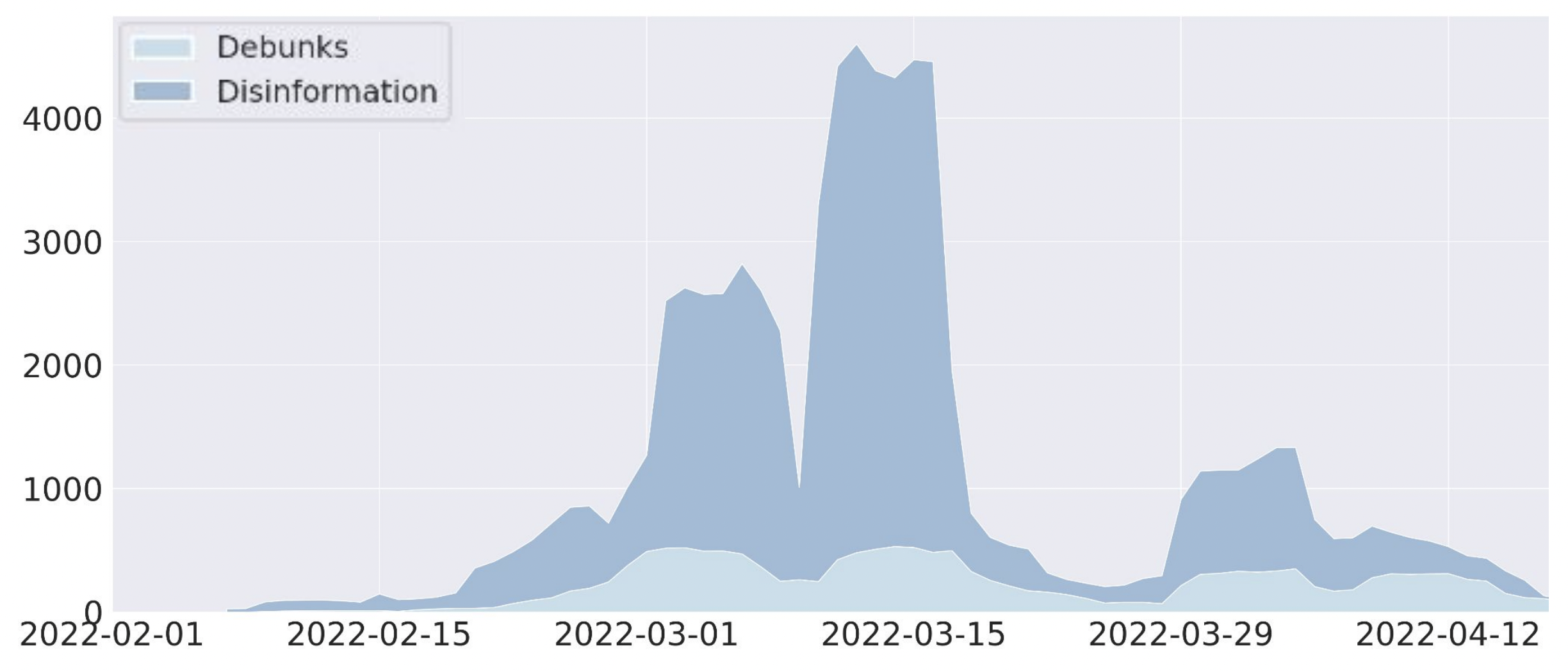}
    \caption{Stacked plot of a rolling 7-day average of the number of disinformation and debunk tweets between 1 February 2022 and 12 April 2022.}
    \label{fig:lineplot}
\end{figure*}

Next, we use academic research access to the Twitter API\footnote{ \url{https://developer.twitter.com/en/docs/twitter-api}} to obtain  16,549 unique tweets containing one of the above debunk URLs and another 62,882 unique tweets sharing one of the disinformation links\footnote{The dataset used for analysis received ethical approval from the University of Sheffield Ethics Board. This paper only discusses analysis and results in aggregate data, without providing examples or information about individual users.}. Retweets are also collected, since we aim to investigate the overall spread of information on Twitter.
Hereafter, the tweets containing debunk links are referred to as ``debunk tweets'' and those containing disinformation links as ``disinformation tweets''. 
Figure \ref{fig:lineplot} shows the stacked plot of a rolling 7-day average curve for the spread of disinformation and debunk tweets. It shows that Ukraine-related disinformation spiked in the first half of March 2022, which consequently lead to an increase in published debunks as it is also reported in EDMO’s Fact-checking Briefs \footnote{\url{https://edmo.eu/fact-checking-briefs}}.

\begin{table}[!tbp]
\small
\centering
\caption{Mean and standard deviation (STD) values of metrics of engagement with disinformation and debunk tweets. * represents a statistical significant difference (p$\le$0.01)}


\scalebox{0.8}{
\begin{tabular}{rcccccc}
\toprule
\multicolumn{1}{l}{}           & \textbf{Followers}~~ & \textbf{Tweets}~~ & \textbf{Retweets}~~ & \textbf{Replies}~~ & \textbf{Likes}~~ & \textbf{Quote count} \\ \midrule
\textbf{Mean - Disinformation} & 6,814                    & 61,331                & 15.0*             & 0.5              & 4.5            & 0.2                  \\
\textbf{Mean - Debunks}        & 21,790*                  & 89,098*               & 1.8               & 0.4              & 2.0            & 0.1                  \\
\textbf{STD - Disinformation}  & 2,04,422                  & 1,22,527               & 312.3             & 34.5             & 631.9          & 13.9                 \\
\textbf{STD - Debunks}         & 4,55,884                  & 1,55,183               & 15.0              & 7.6              & 28.0           & 1.3   \\ \bottomrule           
    
\end{tabular}
}
\label{tab:metrics}
\end{table}

\section{Comparative Analysis of Engagement}
\label{engage}

In order to measure the spread of disinformation and debunks through tweets, we first compare the differences in engagement metrics in terms of mean and standard deviation. Table \ref{tab:metrics} shows the statistics for author's followers, author's tweets, number of retweets, replies, likes and the quote count. We find that the number of retweets, replies, likes and the quote count are comparatively higher for disinformation. However, a $t$-test reveals statistically significant difference ($p\le$0.01) only for the number of retweets, i.e. significantly more Twitter users are retweeting posts containing disinformation URLs than debunk ones. There is also a statistically significant difference ($p\le$0.01) in the number of followers and tweet counts for users sharing debunks as opposed to disinformation. This is as expected since the former are primarily Twitter accounts of fact-checking organisations which naturally have more followers and post more frequently.

\begin{figure}[!t]
    \centering
    \includegraphics[width=7cm,height=3.5cm]{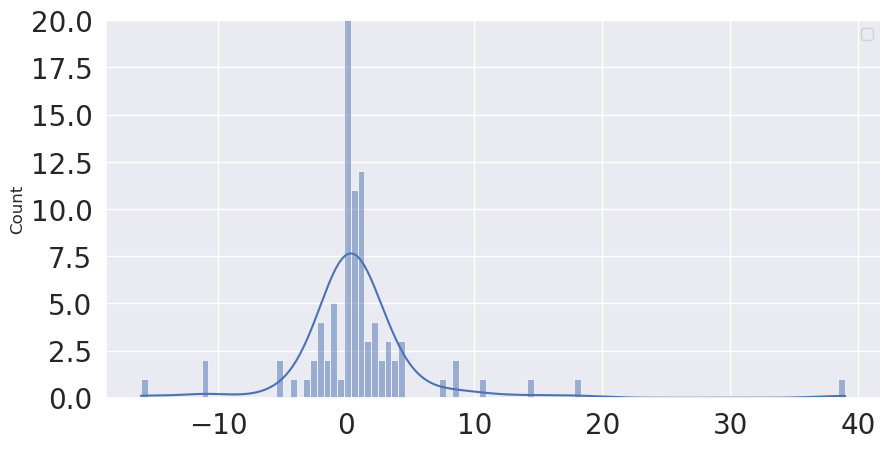}
    \caption{Average difference in days between the date of publication of disinformation tweets and their corresponding debunk article (Fisher-Pearson coefficient of skewness = 3.37).}
    \label{fig:diff_days}
\end{figure}

Figure \ref{fig:diff_days} shows the histogram and kernel density estimate depicting the average number of days between the date of publication of disinformation tweets and their corresponding debunk articles. 
In this, for each debunk we compute  
$
    \sum_{i=1}^{|N|} \mathrm{(\mathit{DoP_{i}} - \mathit{DoP_{debunk}})/ \mathit{|N|}},
$
where $DoP_{debunk}$ is the date of publication of a debunk by the fact-checking organisation, $DoP_{i}$ is the date of publication of a disinformation tweet $i$ and $|N|$ is the total count of disinformation tweets for each debunk.
The data is positively skewed, with a Fisher-Pearson coefficient of skewness of 3.37, suggesting some spread of disinformation even after the publication of the corresponding debunk article (see Section \ref{posthoc}). 

Since EUvsDsinfo debunks explicitly list countries where the disinformation is spreading, these can be compared to the country of the authors of those tweets. 
The latter is derived from the self-declared user location field obtained via the Twitter API\footnote{\url{https://developer.twitter.com/en/docs/twitter-api/data-dictionary/object-model/user}. Where needed,  Geopy Python library (Ref. \url{https://pypi.org/project/geopy/}) is used to extract the country name from the 
information provided by the API.} (when available). We obtain the location information for authors of 51\% of the disinformation-sharing tweets.
Unsurprisingly, the biggest proportion (9\%) comes from cases where EUvsDisinfo has found the disinformation spreading in Ukraine, while the author's self-declared locations are in Russia (Table \ref{tab:spreadcountry}). Another key observation is the global nature of the disinformation, with spread extending significantly beyond Europe.

\begin{table}[!t]
\small
\centering
\caption{Top ten cases with country affected by the disinformation and country of the authors of disinformation tweets.}
\scalebox{0.8}{
\begin{tabular}{ccc}
\toprule
\textbf{Affected country}~~~ & \textbf{Authors' country}~~~ & \textbf{Percentage} 
\\ \midrule
Ukraine                       & Russia         & 9.0        \\
Ukraine                       & Germany        & 7.0        \\
Russia                        & Russia         & 6.0        \\
Russia                        & Germany        & 5.0        \\
Ukraine                       & United States  & 4.0        \\
Ukraine                       & Venezuela      & 4.0        \\
Ukraine                       & Mexico         & 3.0        \\
United States                 & Mexico         & 3.0        \\
\midrule
\multicolumn{2}{c}{Other}                      & 59.0    \\ \bottomrule
\end{tabular}
}
\label{tab:spreadcountry}
\end{table}

\begin{figure}[!tbp]
    \centering
    \includegraphics[width=\textwidth,height=4cm]{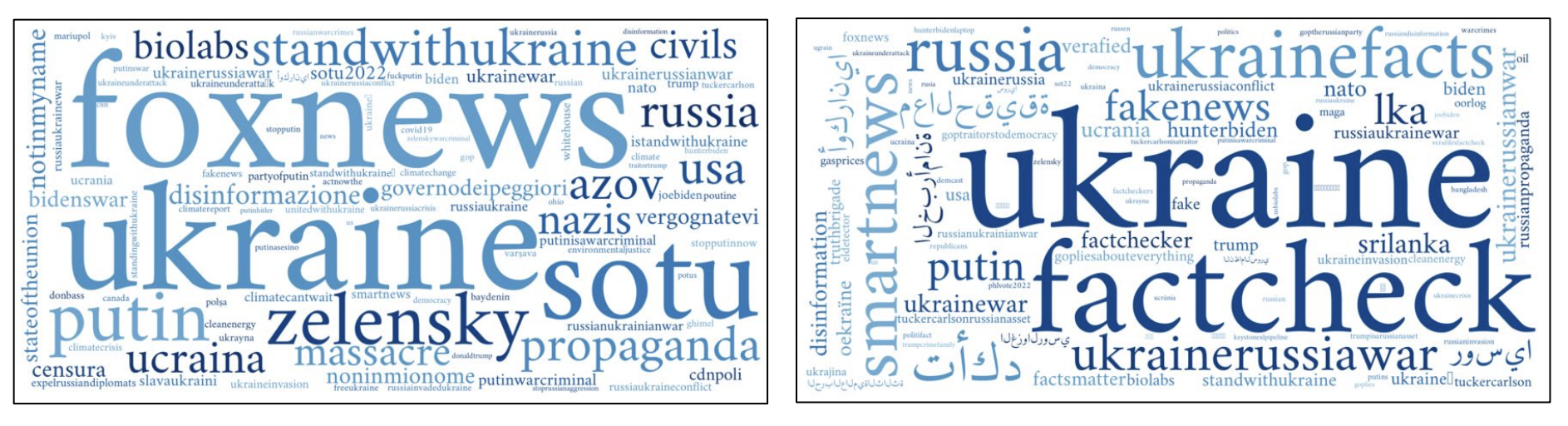}
    \caption{Wordcloud of the most frequent 100 hashtags in disinformation- (left) and debunk-sharing (right) tweets respectively.}
    \label{fig:wordcloud}
\end{figure}

Figure \ref{fig:wordcloud} shows the most frequent 100 hashtags in disinformation-sharing vs debunk-sharing tweets. Unsurprisingly Ukraine dominates both, while {\tt \#FoxNews} is prevalent in tweets sharing disinformation links. This is due to the spread by right-wing American media of a wide-reaching false narrative regarding the presence of U.S.-backed bioweapon labs in Ukraine\footnote{\url{https://www.politifact.com/article/2022/mar/11/russia-china-and-tucker-carlson-lack-evidence-ukra/}}. 

\begin{figure*}[!tbp]
    \centering
    \includegraphics[width=\textwidth]{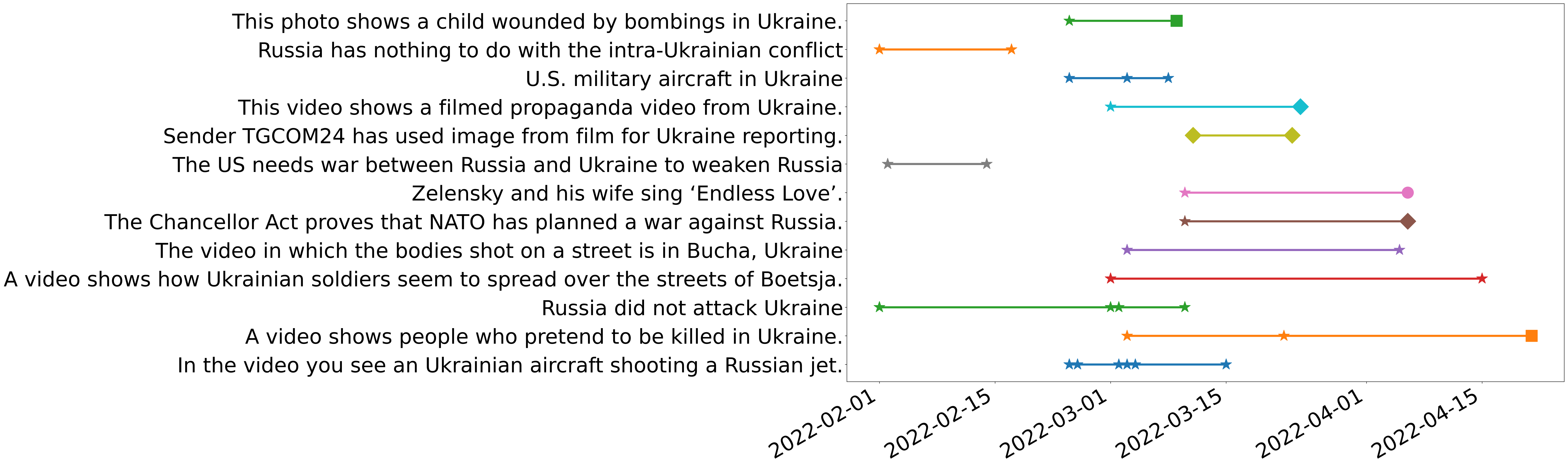}
    \caption{
    Timeline for a sample of Ukraine-related false narratives that have been debunked multiple times. The y-axis states the false narrative and the x-axis represents the date of publication of its debunks.
    The language of debunk articles is denoted by different symbols -- English: $\bigstar$; French: $\blacksquare$; Dutch: $\bullet$; German: $\blacklozenge$ } 
    \label{fig:dupDebunks}
\end{figure*}

We also investigate the presence of identical or highly similar false claims in our dataset that have been debunked multiple times by different fact-checkers. Similar to  \citet{singh2021false}, 
a state-of-the-art semantic search model
\footnote{The multilingual model available at \url{https://huggingface.co/sentence-transformers/paraphrase-multilingual-mpnet-base-v2}, since it performs best according to the leaderboard (Ref. \url{https://www.sbert.net/docs/pretrained_models.html}).} is used for this task. 
Out of the 456 debunks in our dataset (see Section \ref{dataset}), 84 of them (18\%) were found to be highly similar to false narratives that have already been debunked by another fact-checking organisation. Figure \ref{fig:dupDebunks} shows some examples of Ukraine-related false narratives that have been debunked multiple times in different languages. This finding demonstrates significant overlap in effort spent by fact-checking organisations in multiple countries, as well as cost- and time-saving opportunities that could be exploited with the help of translation and cross-publishing of debunks.

\section{Post-hoc Causality Analysis}
\label{posthoc}

We test the bi-directional Granger causality \cite{granger1969investigating} between the disinformation-sharing and debunk-sharing tweets. In other words, we want to investigate 
whether the spread of debunks has a positive impact on reducing the sharing of Ukraine-related disinformation on Twitter. 
Although identifying causation relationships between different information types is not trivial, a Granger causality test can be used to evaluate the predictive causality i.e. if the spread of one information type can be used to predict the spread of another.
In this, we treat the occurrence of disinformation and debunk tweets as two time series variables and then try to find if one variable can be predicted from the other variable’s past values and its own past values. \footnote{The Statsmodel Python library is used to perform the Granger causality test. Ref. \url{https://www.statsmodels.org/}}
First we build a Vector Regression model (VAR) \cite{sims1980macroeconomics}, where a period of three is applied, based on 
the Akaike’s Information Criterion. 
The Augmented-Dicky Fuller test identifies the data as stationary ($p\le$0.01). 
The general equation of VAR model is 

\begin{equation}
    disinfo(t)=\sum_{i=1}^{k} \alpha_{1, i} disinfo(t-i)+\sum_{i=1}^{k} \beta_{1, i} debunk(t-i)+\epsilon_{1}
\end{equation}

\begin{equation}
    debunk(t)=\sum_{i=1}^{k} \alpha_{2, i} debunk(t-i)+\sum_{i=1}^{k} \beta_{2, i} disinfo(t-i)+\epsilon_{2}
\end{equation}

\noindent{where $debunk(t)$ and $disinfo(t)$ refers to count of tweets at time $t$, $k$ is the maximum lag order, $\alpha_{1, i}$ and $\alpha_{2, i}$ are autoregressive coefficients, $\beta_{1, i}$ and $\beta_{2, i}$ are regression coefficients, and $\epsilon_{1}$ and $\epsilon_{2}$ are error terms. }

The experiments find a Granger causality relation, which shows that debunk spread has predictive causality over disinformation spread ($p\le$0.01). In addition, we also observe this weak causation in the opposite direction, i.e. from disinformation to debunks ($p\le$0.01). The significant results in both directions imply that changes in the spread of disinformation may induce changes in the spread of debunks and that the spread debunks may likewise cause changes in the disinformation spread.
This is similar to the findings of previous work for COVID-19 misinformation \cite{burel2020co,burel2021demographics}. 
In order to further understand the weak causation between Ukraine-related disinformation and debunks, we use the VAR model to perform an impulse response analysis and forecast the error variance decomposition for 14 days periods. 

\begin{figure}[!t]
    \centering
    \includegraphics[width=9cm,height=8cm]{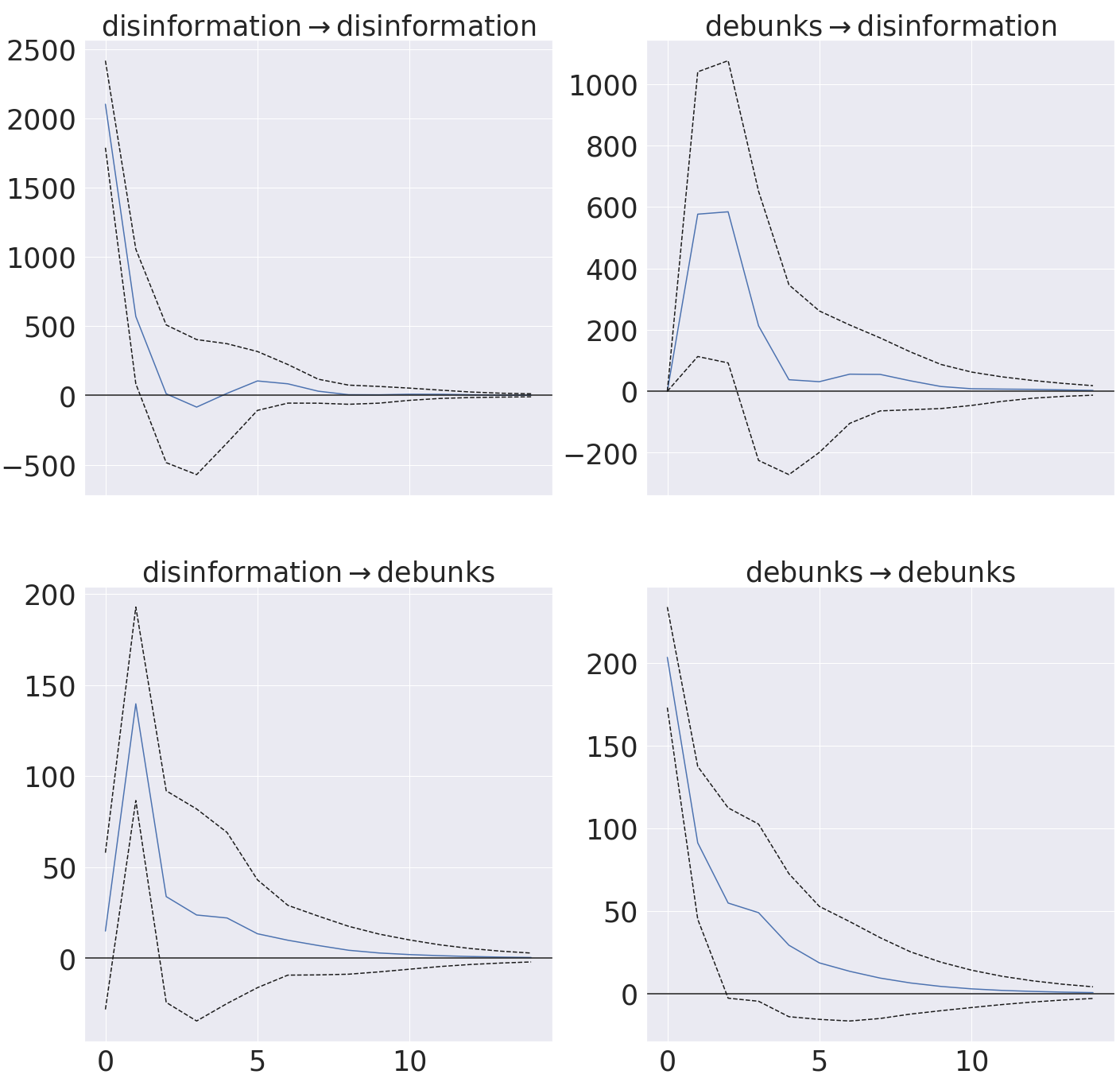}
    \caption{Impulse Response Analysis (x-axis represents 14 days period and y-axis represents effect of shock). Top left and bottom left shows the effect of disinformation shock on disinformation and debunks respectively. Top right and bottom right shows the effect of debunk shock on disinformation and debunks respectively. By default, asymptotic standard errors are presented at the 95\% confidence level.}
    \label{fig:irf}
\end{figure}

Impulse response analysis is used to find the effect of shock in one variable to itself and the other variables in the VAR model.
The prime reason to investigate this is to check if an increase in the spread of debunks triggers a reduction in disinformation on Twitter. 
Figure \ref{fig:irf} shows that an orthogonal shock\footnote{Cholesky decomposition is used for orthogonalisation} from debunks leads to an initial spike in disinformation but there is a downward trend for disinformation afterwards. This suggests that debunks will trigger a reduction in overall disinformation eventually, if not instantly. 
Similarly, an orthogonal shock from disinformation also triggers an initial spike in debunks (Figure \ref{fig:irf}) and an eventual decrease with time, although not instantly. This suggests swift response in debunk publication (mostly from fact-checkers) following a sudden rise in disinformation on social media.
Interestingly, we also notice that the shock in disinformation quickly dies as the impact returns to zero with a sharp decrease on the second day, followed by a small post-shock peak during 4--6 day period which finally converges back to zero between 8--10 day period.

\begin{figure}[!t]
    \centering
    \includegraphics[width=\textwidth,height=4.2cm]{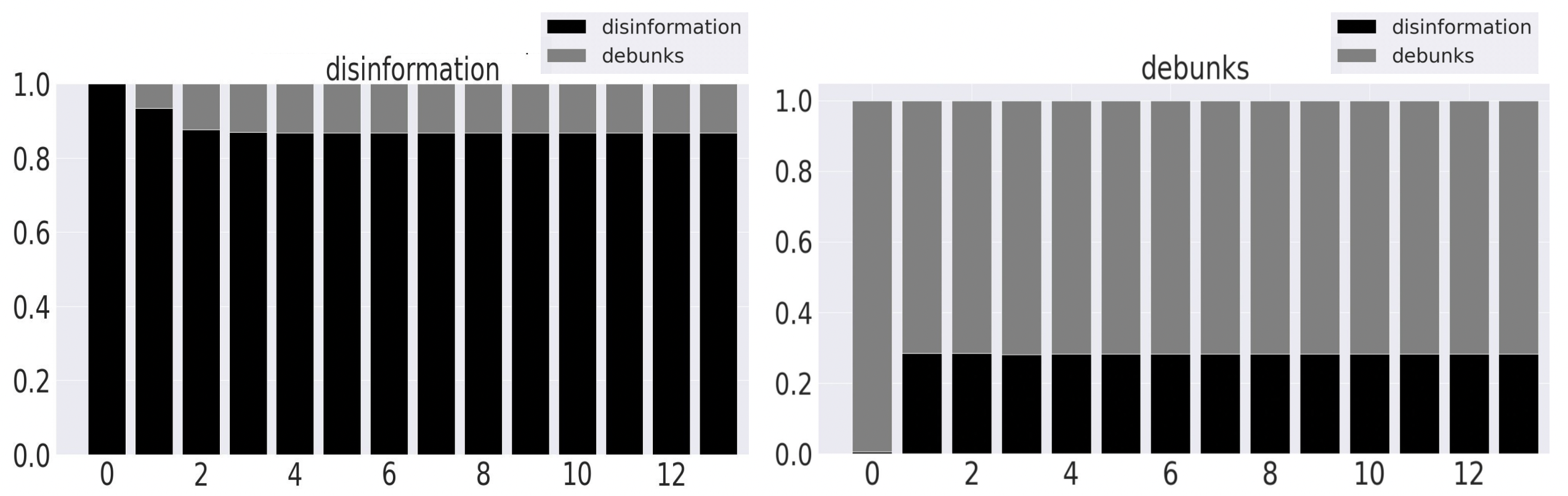}
    \caption{Forecast Error Variance Decomposition (FEVD) plot (x-axis represents a 14-day period and y-axis represents proportion of affect). Left and right represents FEVD for disinformation and debunks respectively.}
    \label{fig:fevd}
\end{figure}

Forecast Error Variance Decomposition (FEVD) helps uncover the proportion of information each variable contributes in predicting a particular variable in the VAR model.
FEVD analysis (Figure \ref{fig:fevd}) reveals substantial predictive dependencies between Ukraine-related disinformation and debunks. Similar to what we find in impulse response analysis, FEVD results show that debunks directly affect disinformation by around 15\% by the end of the 14-day period. We also observe that debunks affect the spread of disinformation after an initial delay by a day, after which it rises and becomes constant following the third day. 
On the other hand, we also find that the spread of debunks is also affected by disinformation by around 30\%. The results also show that the impact of disinformation on debunks is delayed initially for a day. In other words, this implies that the spread of debunks is not dependent on how Ukraine-related disinformation spreads initially. 
In summary, our experimental analysis confirms that the spread of debunking tweets does have a positive impact on reducing Ukraine-related disinformation on Twitter.

\section{Topical Analysis of Ukraine-related Disinformation}
\label{topic}

This section investigates the main topics in disinformation and study the engagement around them over time. The debunked claim statements are clustered by applying K-means to embeddings from the semantic search model\footnote{We use the BERTTopic \cite{grootendorst2022bertopic} Python library for clustering and MPNet \cite{song2020mpnet} as the transformer model. Ref. \url{https://huggingface.co/sentence-transformers/all-mpnet-base-v2}}. The number of clusters is kept at six using the Elbow method and silhouette coefficient score. The model is run for a maximum of 300 iterations with K-means++ used as a method of initialisation. The clustering is applied on debunked claim statements and not on tweets itself. See Appendix \ref{appendix:datacollect} for details on how we collect the debunked claim statements.

\begin{table*}[!tbp]
\centering
\small
\caption{Top ten words and count of disinformation tweets in each topic cluster. Order of word depicts its importance from left to right.}

\scalebox{0.8}{
\begin{tabular}{lc}
\toprule
\textbf{Topic clusters}                                                                       & \textbf{Count} \\ \midrule
0\_ukraine\_ukrainian\_russia\_kyiv\_neo\_coup\_nazis\_war\_crimea\_weapons                   & 39,761          \\
1\_poland\_nato\_polish\_alliance\_west\_security\_countries\_western\_europe\_russian        & 10,225          \\
2\_putin\_biden\_know\_think\_vladimir\_lee\_answer\_prices\_says\_oil                        & 4,194           \\
3\_video\_shows\_ukraine\_ukrainian\_proof\_jet\_marcos\_shot\_soldiers\_fighter              & 3,793           \\
4\_biolabs\_ukraine\_financed\_state\_biological\_labs\_military\_biden\_victoria\_vaccinated & 3,132           \\
5\_trump\_russia\_bucha\_massacre\_billions\_evidence\_west\_100\_planted\_united             & 1,777  \\ \bottomrule        
\end{tabular}
}

\label{tab:topinc}
\end{table*}

The class-based TF-IDF \cite{grootendorst2022bertopic} is used to find top words in debunked claim statements in each of the clusters (Table \ref{tab:topinc}). Each cluster has distinct words that separate it from the other five. In order to verify the separation  between the clusters, we also plot a heatmap of topic similarity (see Appendix \ref{appendix:Heatmap}). The results show that except clusters one and two, most of the clusters are distinct in terms of the topics they cover. For instance, cluster four encompasses wide-spread conspiracies related to the U.S.-backed bioweapon labs in Ukraine and the US planning to send infected migratory birds to infect Russia\footnote{\url{https://euvsdisinfo.eu/report/the-us-plans-to-send-infected-migratory-birds-to-infect-russia}}. Another cluster (one), includes the ongoing false narrative about the NATO country alliance being the real threat to Russia\footnote{\url{https://www.politifact.com/factchecks/2022/feb/28/candace-owens/fact-checking-claims-nato-us-broke-agreement-again/}}. 
Table \ref{tab:topinc} also shows the count of corresponding disinformation tweets (and retweets) in each cluster, identifying that most of the tweets belong to cluster zero and one. 

\begin{figure*}[!tbp]
    \centering
    \includegraphics[width=13cm,height=5.4cm]{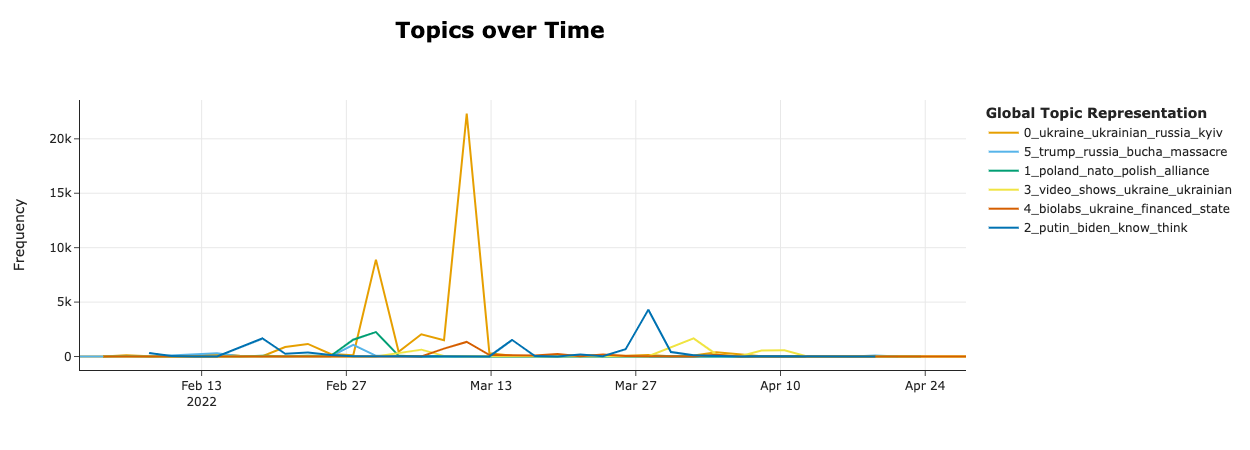}
    \caption{Temporal spread of disinformation tweets in each topic cluster over time. Legend shows top five words of each cluster from Table \ref{tab:topinc}.}
    \label{fig:timelineplot}
\end{figure*}


Next, we look at the temporal distribution of the disinformation tweets for each topic cluster. Figure \ref{fig:timelineplot} illustrates the line plot for topic prevalence between February and April 2022. For instance, cluster zero includes false claims related to Russia attacking Ukraine, Kyiv, neo-nazism, etc. and has two dominant peaks, one in the first week of March and another one in the second week of March (the tallest one at 10 March 2022). There is also an uptick in February suggesting that the disinformation narratives started spreading even before 24 February and then spiked later in March. Similar results were found in the EDMO’s Fact-checking Briefs for February\footnote{\url{https://edmo.eu/fact-checking-briefs}} where they noticed a sudden increase in posts about growing tensions between Russia and Ukraine. 

Cluster one (disinformation related to NATO and western countries) spiked in the first week of March. It includes a dominant false narrative about NATO attacking countries illegitimately\footnote{\url{https://euvsdisinfo.eu/report/nato-is-not-a-defensive-alliance-it-attacks-countries-illegitimately}}. 

Cluster two has multiple spikes: one in mid February; another one in mid of March; and the biggest one -- at the end of March. It comprises false narratives involving Joe Biden, Vladimir Putin and Russian oil, e.g. that Biden's cancellation of the Keystone pipeline ``dramatically increased Americans'' dependence on Russian oil.

Cluster three comprises of videos spreading disinformation and their distribution is fairly stable, with only a slight increase in the first and last week of March.

Cluster four contains conspiracies, such as an alleged presence of US biological labs in Ukraine\footnote{\url{https://www.bbc.co.uk/news/60711705}} and release of infected migratory birds to infect Russia\footnote{\url{https://euvsdisinfo.eu/report/the-us-plans-to-send-infected-migratory-birds-to-infect-russia}}. Figure \ref{fig:timelineplot} shows that these type of conspiratorial narratives spiked during the second week of March. This is also coherent with the findings of EDMO’s Fact-checking Briefs for March 2022.

Cluster five contains disinformation related to Trump and the Bucha massacre and has comparatively time-limited span, with only a small peak at the end of February 2022.


\section{Limitations and Future Work}
Our work should be seen in the light of the following limitations. First, as described in Section \ref{dataset}, the study uses only tweets which contain explicit links to known disinformation or debunk articles. While this makes the dataset highly accurate and does not require additional human annotation, 
it also means that tweets that spread false claims or debunk them without citing a reference link could not be included.
Second, this paper only discusses results on aggregate data, without looking at whether the tweets are from real or bot accounts.
Lastly, user data, such as their country, is dependent on self-declared information in the user profiles, which is missing for many tweets. Nevertheless, the sample size is sufficiently large and robust to yield useful insights.

In future, we want to analyse the spread of disinformation and debunks before and after the start of the Russia-Ukraine war. We might have different answers for the research questions raised in the paper, which would potentially provide some insights into how an emergency event changes the spreading paradigms of disinformation and debunks. We also want to find ways to automatically detect disinformation tweets which don't explicitly mention links or where the disinformation links mentioned are different from the ones present in our dataset. Lastly, the Granger causality test deals with linear relationship. 
Hence, in future, we plan to experiment with other non-linear tests like Hiemstra and Jones non-linear Granger causality \cite{hiemstra1994testing} and Convergent Cross Mapping test \cite{tsonis2018convergent}.

\section{Conclusion}
\label{conclu}

This study carried out a comparative analysis of the spread of Ukraine-related false claims and debunks on Twitter between February and April 2022. In particular, our comparative engagement analysis found that tweets spreading disinformation are shared and retweeted significantly more as compared to those containing debunks. With respect to debunks, we also established that around 18\% are focused on false claims for which debunks have already been posted in a different country or language. This finding is particularly important, as it points out to two opportunities going forward. Firstly, since many platforms, such as Facebook, already offer machine translation tools to their users, they could us that technology themselves to translate and match debunks automatically, so a false narrative spreading in one language can be flagged as false, based on an authoritative fact-check in another language. Secondly, fact-checkers themselves can benefit from using cross-lingual search and machine translation technologies to find such debunks, which they can then cite as a source or re-publish in translation and thus reduce the time elapsed between a false narrative starting to spread widely online and the time their debunk is published. 

Another key finding is that the publication of debunks does ultimately lead to limiting the spread of Ukraine-related disinformation, albeit not immediately. 
In addition, FEVD results show substantial predictive dependencies between the spread disinformation and debunk tweets.
Lastly, our data-driven analysis uncovered also the dominant themes in Ukraine-related disinformation and their temporal intensity.
In conclusion, these findings have immediate relevance for a wide range of stakeholders, including digital platforms, fact-checkers, and online information users. The dataset used for analysis is available at {\tt \url{https://doi.org/10.5281/zenodo.6992686}}.



\subsubsection{Acknowledgements.} This research has been partially supported by  a European Union – Horizon 2020 Program, grant no. 825297 (WeVerify), the European Union – Horizon 2020 Program under the scheme “INFRAIA-01-2018-2019 – Integrating Activities for Advanced Communities” and Grant Agreement n.871042 (“SoBigData++: European Integrated Infrastructure for Social Mining and Big Data Analytics” (http://www.sobigdata.eu)).

\bibliographystyle{splncsnat}
\bibliography{bibliography}

\appendix

\section{Appendix}
\subsection{Data Collection}
\label{appendix:datacollect}

\begin{figure*}[!htbp]
    \centering
    \includegraphics[width=12cm,height=7cm]{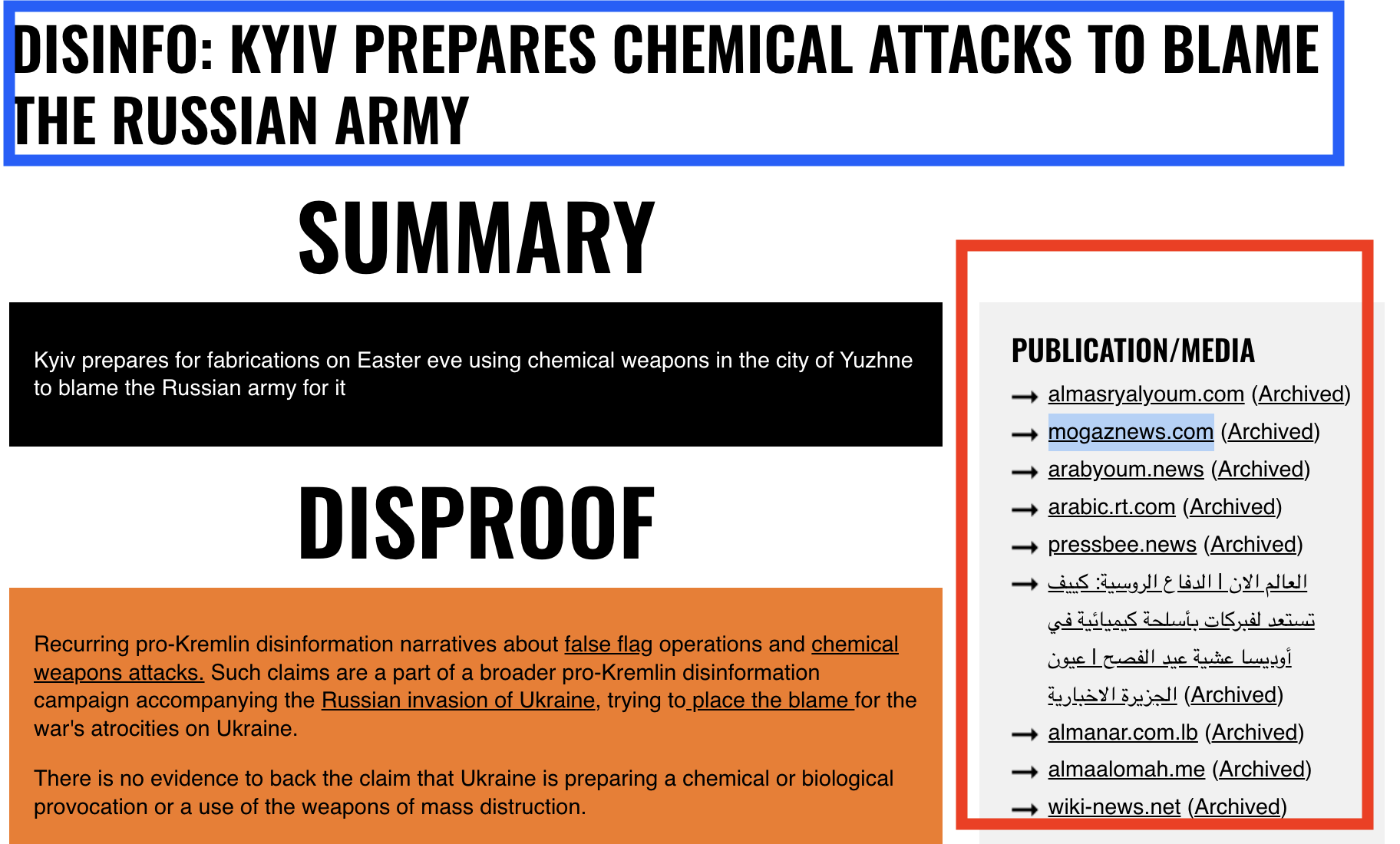}
    \caption{Screenshot of one of the EUvsDsinfo debunks. Section enclosed in the \textcolor{red}{red} box contains disinformation links and the \textcolor{blue}{blue} box represents the debunked claim statement.}
    \label{fig:euvsdisinfo}
\end{figure*}

As described in Section \ref{dataset}, we collect Ukraine-related
debunks from EUvsDsinfo and  ClaimReview. In order to collect the disinformation links, 1) the debunks indexed in ClaimReview schema has the {\tt itemReviewed}\footnote{\url{https://schema.org/ClaimReview}} object which includes disinformation links that are being debunked by fact-checking organisation and debunked claim statement is present in  {\tt claimReviewed} object; 2) the debunks on EUvsDsinfo explicitly mention disinformation links on their website. Figure \ref{fig:euvsdisinfo} shows the screenshot of one of the EUvsDsinfo debunks. The section enclosed in the red box contains disinformation links and the blue box represents the debunked claim statement.

\subsection{Heatmap}
\label{appendix:Heatmap}

\begin{figure}[!t]
    \centering
    \includegraphics[width=6cm,height=6cm]{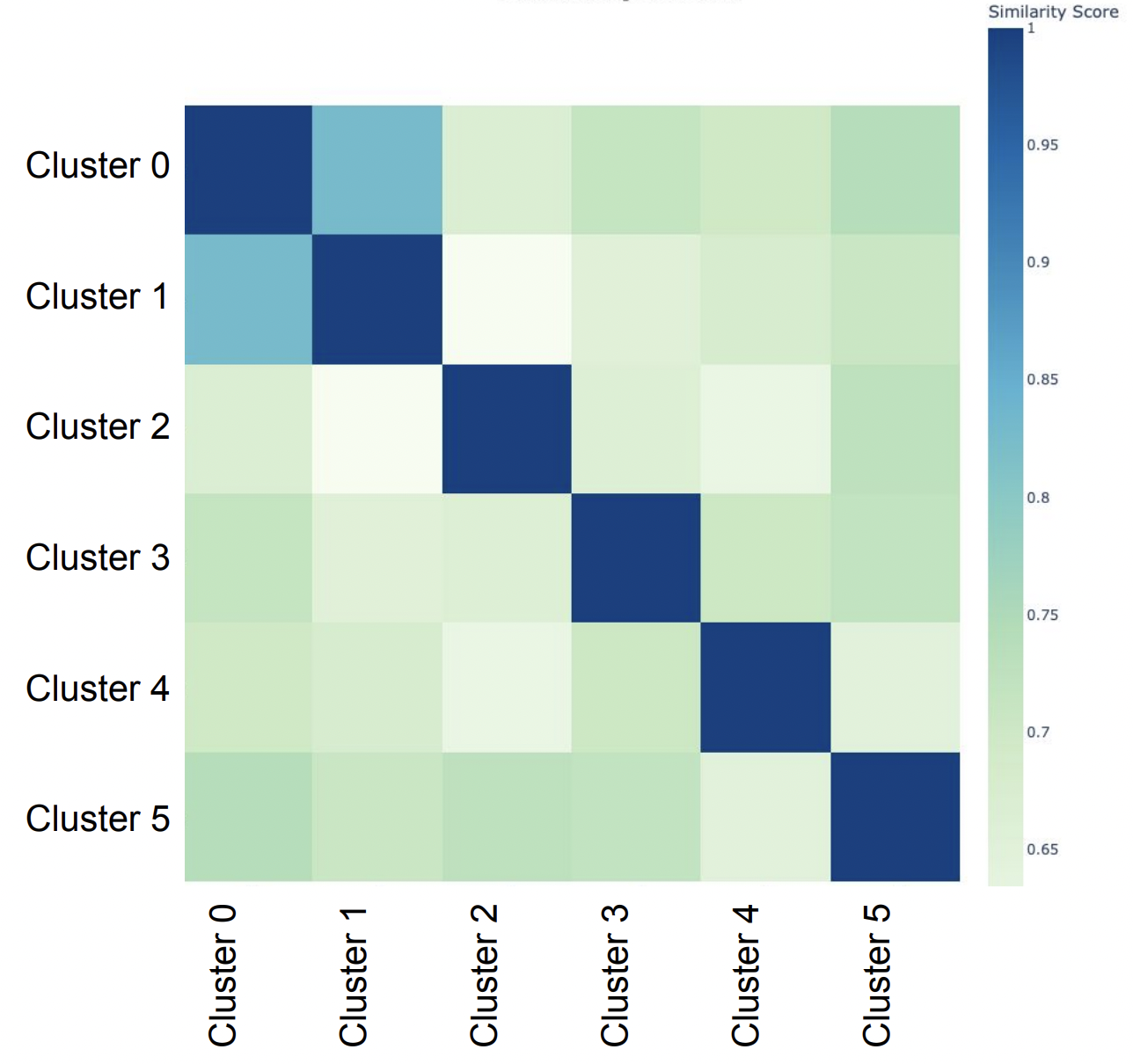}
    \caption{Heatmap for topic cluster similarity. The description of clusters can be found in Section \ref{topic}.}
    \label{fig:corr}
\end{figure}

Figure \ref{fig:corr} illustrates the heatmap of cluster similarity. The results show that except clusters one and two, most of the clusters are distinct in terms of the topics they cover. This indicates reasonable separation between the clusters found in Section \ref{topic}.

\end{document}